# The Influence of Mean Motion Resonances on the Outer Kuiper Belt: Does the Outer Kuiper Belt have a Future?

Fred A Franklin and Paul R Soper
Harvard-Smithsonian Center for Astrophysics

It often seems as though papers bearing titles in the form of a question end with ambiguous answers.   Here the situation is different: the outer Kuiper belt does have a definite future, although one of uncertain duration.   Simulations provide two distinct, compelling reasons.  First, mean motion resonances in the outer belt [i.e., beyond the 1/2 resonance at 47.76 AU] are amazingly "sticky": in almost all cases bodies captured in them from regions closer to Neptune during that planet's outward migration remain trapped for 4.6 byr.  Most captured orbits are chaotic and so will eventually escape, but there is no reason to believe that all outer belt resonances will empty in the near future.   Second, in determining capture probabilities for various resonances, we find that the first order 1/2 resonance is quite efficient, but, in clear contrast to higher order ones in the outer belt, nearly half of its victims have escaped during 4.6 byr.  These bodies typically remain in unstable orbits in the outer belt, usually for several hundred million years before being expelled well beyond 100 AU.  Other inner belt resonances, although capturing less efficiently than 1/2, behave in the same fashion.  Thus the outer belt has a two component, long-term population: one with members lying semi-permanently in one of about 6 resonances and another that is far more temporary, but whose source provides a continuous and ready resupply. The resulting eccentricity distribution of this combined population closely matches the observed one, as Hahn and Malhotra have made clear but the inclination distribution is a better approximation to it than is often claimed, a fact that probably should not be ignored when considering an origin for what has been labeled "the scattered disk".

The ability of Neptune's migration to populate the outer Kuiper Belt from bodies initially closer to the planet was carefully developed by Hahn and Malhotra [H & M] (2005).  Our efforts here, while also in the way of confirming their conclusions, focus on the characteristics and long-term influence of mean motion resonances [mmrs], especially in the outer belt, a > the 1/2 mmr now at 47.8 AU, and to some extent, in the inner one, a < a(1/2) as well. [Here we refer to resonances by the mean motion ratio of the outer body, a KBO or TNO, to the inner, Neptune.]  We shall pursue the hypothesis that an inner belt of primordial bodies, plus Neptune's parameterized migration, is all that is required to explain the presence of the current outer belt, to maintain it in the future and to account for the distribution of important orbital elements of its residents as well.  The second point is a principal reason for this paper. Evaluating the potential significance of mmrs requires knowing 1) capture probabilities, 2) characteristic escape times following capture and 3) the range of orbital elements after escapes.   Mmrs of first-order (e.g. 1/2) through at least seventh have an impact on the dynamical history of both the inner and belts.

To summarize two useful results obtained from integrations of sets of 500 bodies: 1) certain mmrs play a neat game of "catch and release"—catching a fraction of primordial bodies during multi-million year migrations and then releasing some of them, while retaining others, at larger than their initial semimajor axes, a, at higher eccentricities, e, and inclinations, i, over times extending up to the present and seemingly well into the future as well.  Thus we find an automatic and continuous means to replenish the population of the region out to 100 AU, a limit we impose here because of reasonable observational coverage and where the accuracy of our integrations remains high. The 1/2 mmr captures effectively, but objects trapped there generally have very chaotic orbits and nearly half of them escape from it within 4.6 byr. Escapers from 1/2 and some other mmrs with a < a(1/2) are eventually removed from the outer belt over widely varying time scales, averaging near 300 myr, either by being driven to still larger distances, a > 100 AU, or by colliding the major planets.
Resonances lying at a > a(1/2), but of higher order, e.g., 1/3, 2/5 and 3/7, do capture bodies during migrations, though less efficiently than 1/2, and, while most resulting orbits are chaotic, only two cases of escape from a single 6$^{th}$ order one have occurred up to 4.6 byr. Thus the two component population of the outer belt consists of bodies of long-term residency in mmrs like the three just mentioned, mixed with temporary, unstable objects that have escaped from 1/2 and other mmrs with a < a(1/2).

The second result is related to the first: as discussed by H & M, capture at assorted mmrs during Neptune's migration, sometimes followed by escape, readily explains the eccentricity distribution that rises with a for a > a(1/2). But we now note here that capture and escape may also account for much of the elevated inclination distribution among the escaped bodies that is essentially independent of semimajor axis. This especially is true for the five bodies that ventured beyond the (arbitrary) 100 AU limit and then returned to the outer belt. We can begin to make a fair case that no special mechanism is required because perturbations, often involving several mmrs acting sequentially, can increase <i>'s to form what has been labeled "the scattered disk".

To support these claims the next paragraphs present quantitative material dealing with capture probabilities, escape times and the time dependent distribution of orbital elements. Table II examines the first of these three for a case in which Neptune migrates outward by 7 AU, ending at its current semimajor axis of 30.09 AU, over a smooth e-folding time scale, T(m), of 10 myr. All integrations include the other major planets, moving in their present orbits. The figure of 7 AU [cf Luu and Jewitt, 2004; H & M, 2005] is a likely, but near maximum value. It has, given their mass difference of a factor of 18, a measure of consistency with Jupiter's observationally based inward migration of about 0.45 AU (Franklin et al, 2004). Table II relies on 500 massless bodies initially having the elements: 38 < a(o) < 52 AU; 0.04 < e(o) and sin i(o) < 0.15. Our adoption of elements as high as these leans on the conclusion of H & M: a "cold" inner belt population at the time of migration cannot satisfy the currently observed distribution of its e's and i's.

As Table I clarifies, this semimajor axis range allows a search for possible captures into mmrs lying just beyond initial values of a(1/2) when Neptune travels outward by 7 AU. Table II lists capture probabilities, P©, for the three strong mmrs; another, 1/3, has too few examples hereso we undertook a simulation with extended limits [cf Table III]. We define P© as the number of captures [that show a clear libration] divided by the number of available bodies in the appropriate a and e ranges. All P©s are corrected for earlier sweeping, e.g., 2/5 scans part of the range before 1/2, hence some of the bodies captured into 2/5 are not available for 1/2. No appreciable dependence of the P©s on i was apparent for 2.3 < i(o) < 8.6 deg.
Table II shows that captures into 1/2 are numerous, but, lying near 20 to 25%, not inevitable, while the higher order mmrs 2/5 and 3/7 remain effective at a reduced but not insignificant level. Thus the rather expected result: a single sweep by several mmrs will not seriously depopulate an initial semimajor axis range. What is in clear contrast with the behavior at 1/2 is the ability of the higher order mmrs to capture bodies with larger initial eccentricities so effectively. Some insight into this behavior, noted by Chiang and Jordan (2002), is included in H & M. The inability of scanning by mmrs to remove a large fraction of bodies, plus the complete absence of real objects with e < 0.10 in the region a > a(1/2) and their scarcity as high as e = 0.20, [cf Figs. 3 and 4] combine to insist that very few bodies were ever there. Thus the claim that the present population of the outer belt owes its origin to the transfer of objects from the inner one seems very sound.

The P© for 1/3 in Table II is uncertain, which one might expect to be a consequence of its smaller collecting range of only ~ 4 AU [cf Table I] in the first simulation   To improve its accuracy and to check on P©s with e(o)s up to 0.2, we introduce for this purpose only the two cases summarized in Table III, where the semimajor axis range over which captures can occur extends now to include much of the outer belt, 48.0 < a(o) < 62.6 AU.  These results, again both based on 500 initial objects, strengthen some of the numbers in Table II and also argue that 1) capture at 2/5 is either curiously effective or 1/3 ineffective, 2) there is little change in the P©s for e(o)s as high as 0.2 and 3) captures at 5$^{th}$ and even occasionally at 7$^{th}$ order mmr, as listed also in Table II, do occur.

The influence of the duration and magnitude of Neptune's migration is the reason for Tables IV and V, where the former uses the same orbital elements as the case in Table II but a T(m) of only 3 myr and the latter corresponds to a migration of 8 AU, but over 10 myr.   Taken together, Tables II - V, despite certain minor differences, reinforce the preliminary conclusions sketched earlier, viz.:
a) The 1/2 mmr will capture some 20-25% of all bodies it encounters in the  range 0.04 < e(o),sini(o) < 0.15, but does so most effectively for e(0) <  0.1.
b) The 3$^{rd}$ order 2/5 mmr captures at close to 5%, but the 2$^{nd}$ order 1/3 mmr is, for all case examined, markedly less efficient than 2/5.
c) mmrs of order 2 and higher capture more effectively for e(o) > 0.1 and this behavior is increasingly pronounced as the order rises.

Concerning point a): capture probability for the 1/2 mmr is a topic covered by Chiang and Jordan [2002].  Also with a T(m) = 10$^7$  yr, they obtained P© = 53% for the case when the initial eccentricity distribution lies in the low range: 0 < e(o) < 0.05, showing in addition that it drops steeply as T(m) decreases, falling to 0 at T(m) = 10$^5$ yr.  Our values seem quite consistent with theirs [especially for the ones in Table II that also adopt T(m) = 10$^7$  yr] inasmuch as ours, applying to the larger ranges of eccentricity that are divided into two parts: 0.04 - 0.10 and 0.10 - 0.15, yield P©s of 31 and 12% respectively.  Results in Table IV also favor agreement as P© for the low e(o) case declines from 31 to 23% as T(m)drops from 10 to 3 myr.  How the larger e(o) case depends on T(m) needs a closer look.
Current observations processed at the Minor Planet Center [MPC], as of June, 2012 and summarized in Table VI, generally agree with conclusion b).  They indicate that the population of KBOs in 2/5 is ~1.5 times greater than in 1/3, while the 2/5 and 3/7 ratio is close to 2 to 1.  Some caution is called for as the observed numbers of bodies assumed to lie in a mmr correspond just to those with semimajor axes lying within +/- 0.5 AU of a resonance, a characteristic value of the present uncertainty in a as evaluated by Gareth Williams of the MPC.

To sharpen the predicted relative populations, we turn first to the two cases compiled in Table III.  Corrections to the tabular entries, chiefly involving the 2.1 times larger collecting area for 1/3, yield a ratio of the number of objects in 2/5 to 1/3 of 1.3 to 1.  Better, or greater numbers for the 2/5 to 3/7 ratio come from Table II and applying appropriate corrections for collecting gives a 2 to 1 population ratio for 2/5 relative to 3/7.

The somewhat surprising result, b) and c) as well, suggest the value of checking details of mmrs in the region a < a(1/2).  Resonances are more densely spaced there, often leading to a degree of overlap, but the P©s in Tables VII and VIII are ones that apply when it's clear that only one mmr is involved. Cases of escape from one mmr followed by a capture into another are therefore excluded. These two tables argue that the 2nd order 3/5 mmr nearly rivals 1/2 in capture efficiency—far more so than its 1/3 relative—while jointly with other tables, they suggest the P©s drop by a factor of roughly two between successive orders.

The second topic of interest centers on stability after capture: will bodies escape from various mmrs and if so, over what time scales? Table IX, drawn from the same data set as Table II, begins to examine this question. Of the 69 captures into 1/2, 31 have escaped over times as short as $7.7 \times 10^7$ yr up to the "present" at $4.6 \times 10^9$ yr. The following table lists the number of bodies that have escaped from this mmr in three essentially equal time periods of 1.5 byr.

| Time Interval [in billion yrs] | 0 - 1.5 | 1.5 - 3.0 | 3.0 - 4.6 |
|---|---|---|---|
| Number of Escapers | 14 | 7 | 10 |

[Recall that for these data, $T(m) = 10^7$ yr and all times are measured from the start of Neptune's migration.] Escapers are about evenly divided between e(o)s greater and less than 0.1. In view of the very high frequency of chaotic orbits at 1/2, we can expect additional escapes so that the outer belt, for well into the future, has all the prospects for a built-in, continuing existence. Despite the presence of many chaotic orbits, particularly at 2/5, no escape from this resonance has occurred in 4.6 byr and only two cases of escape from any outer belt mmrs were noted, both from the 5/11 mmr [from a total of 6 captured there] at 2.85 and 3.16 byr. As for the regular orbits, all are of low inclination, with i(max) < 8 deg. and with modest e's of 0.2 to 0.35, though regular orbits at 1/3 attain e's up to 0.45. A discussion of the source of chaos in some outer belt mmrs, especially 1/2, is a topic we turn to in another paper [Franklin and Soper, 2012].

This degree of permanence in mmrs of the outer belt, however, does not obtain in the inner one as Table VII indicates for the case of four examples. The difference in stability on opposite sides of 1/2 is puzzling. Orbital behavior in the inner belt for this simulation is more complex than in the outer because of the higher density of mmrs that sequentially sweep through its [assumed] ambient population on their way to expelling bodies into the initially empty region beyond 1/2. The mmr listed in Table VII join 1/2 in ejecting bodies into the outer belt. The escapes begin quickly, as early as $4 \times 10^7$ yr and continue to $4.6 \times 10^9$ yr, by which time 4 of the 12 captured in 3/5 and 6 of 9 in 4/7 have been ejected.

After a consideration of capture and subsequent escape, two other concerns are: how long do escapers typically linger in the outer belt and what is the range of their orbital elements while there? The first part of this query is easily answered: of the 500 bodies forming the basis of Table II, 156 were transferred out of the inner belt within $4.6 \times 10^9$ yr. Of these 43 were captured into outer belt mmr [with only the two late escapers from 5/11], 9 moved inwards and never reached the outer belt and the remainder, 104, spent some time, averaging $\sim 3 \times 10^8$ yr, in the range 48 < a < 100 AU. Most moved eventually to still greater distances though occasionally revisiting the outer belt. When so, five showed extreme elements, including inclinations of 30 to 65 deg.

We approach an answer to the second part of the question in two ways. Ideally we'd like to plot the distribution and range of elements of bodies in the outer belt as seen today. But because the typical residency time is $\sim 3 \times 10^8$ yr, only ten non-resonant objects remain in a simulation. Most of these, 8 of 10, were fairly recent escapees from 1/2. More to point are Figs. 1 and 2 that provide a sort of cumulative picture, giving time averaged a, e and i values for all bodies that have spent time in the outer belt after Neptune's migration ceased. They draw upon the two classes of bodies: the 41 lasting captures into 6 mmrs located there [4-pointed open stars], while crosses mark the 104 more temporary members. Figure 1 matches well the rise in e vs a shown by the observed distribution in Fig. 3, but Fig. 2 has some special interest as it argues that objects, after escaping from one or more mmr, can acquire i's that are well above the initial range of $2.3 < i(o) < 8.6$ deg.

The data in Fig. 2 are replotted as a histogram in 5 deg. intervals as Fig. 5, providing the frequency distribution of <i> when the time average of their individual a's placed them in the outer belt. The comparison with the observed i's in Fig. 6 [whose numbers have been reduced by a factor of 145/158 so as to be identical with those of Fig. 5] shows only limited agreement, but, together with the role of pumping by secular resonances [Nagasawa and Ida, 2000], places some doubt on the need for other explanations to account for the scattered disk values of the i's.

Although we have not considered the region a > 100 AU, still the role of mmrs that can, during Neptune's migration, capture bodies with a(o) as small as 40 AU [or even less], later followed in some cases by escape, would seem to provide another possibility for the origin of large objects like (90377) Sedna in addition to the one proposed by Kenyon and Bromley (2004). Sedna now lies at a = 537 AU with e = 0.86, i = 12 degrees, elements that an extrapolation of the behavior of bodies shown in Figs. 1 or 3 might be expected to fulfill.

Final Remarks

Our aim in this paper centers on starting a numerical examination of mean motion resonances, mainly at and exterior to the 1/2 mmr, now at 47.8 AU, during and following Neptune's migration. We find a clear, though qualitative difference between 1/2 [together with others closer to Neptune] and those beyond it in the sense that outer belt mmrs, a > a(1/2), retain nearly all bodies captured during the planet's migration for times longer than 4.6 byr. By contrast, 1/2 itself and others with a < a(1/2) have released about half of those captured by that time. Most of these move in unstable, eccentric and inclined orbits in and beyond the outer belt for several hundred million years.

Since a majority are likely to be escaped members of the 1/2 mmr, they might be expected to bear some resemblance to bodies lying there now. The outer belt, 48 < a < 100 AU, is therefore the home of long-lived captured bodies lying in the local mmrs and also the more temporary domain of objects expelled from inner belt mmrs. The resulting orbital distribution is a very good [for eccentricities] to an only fair [for inclinations] representation of the observed ones. Our result showing the permanence of bodies in the outer belt mmrs means that the eccentricity pumping produced by resonant capture for the temporary objects there is achieved only when these objects lie in mmrs of the inner belt—thus those equal to or less than a(1/2). This in turn implies that the eccentricity increase with semimajor axis shown in Figs. 1 and 3 must have contributions from other gravitational perturbations. Put another way, as Fig. 7 makes clear, migration and capture will increase e, even when a mmr moves by ~13 AU, by no more than ~0.3 while Figs. 1 and 3 show values a factor of 2 or more higher.

We would like once again to thank Matt Holman for his symplectic integration program that PRS has modified to include planetary migration.

Figure Captions

Fig. 1: Crosses mark mean e's and a's of 104 [of 500] bodies driven into the outer belt after escapes [overwhelmingly] from resonances lying initially at or below a(1/2). Four-pointed open stars denote 41 other bodies that were captured and retained in the 6 indicated outer belt resonances for 4.6 byr. Circles mark escapers from 1/2.

Fig. 2: Mean inclinations for the bodies shown in Fig. 1. Values here and in Fig. 1 are average values occurring during ~ 300 myr, a typical time spent in the outer belt.

Fig. 3: Eccentricities of all known [June, 2012] bodies in the outer Kuiper belt, i.e., those beyond the 1/2 resonance at 47.8 AU out to 100 AU.

Fig. 4: A companion to Fig. 3: observed inclinations of the 158 outer belt KBOs.

Fig. 5: Histogram in 5 degree inclination intervals of all bodies shown in Fig. 2.

Fig. 6: A similar histogram of the observed objects in Fig. 4, slightly rescaled so as to match the same number of bodies as in Fig. 5.

Fig. 7: Eccentricity increase of bodies captured into inner belt resonances [crosses] and outer ones [squares] up to fifth order during Neptune's 10 myr and 7 AU migration.

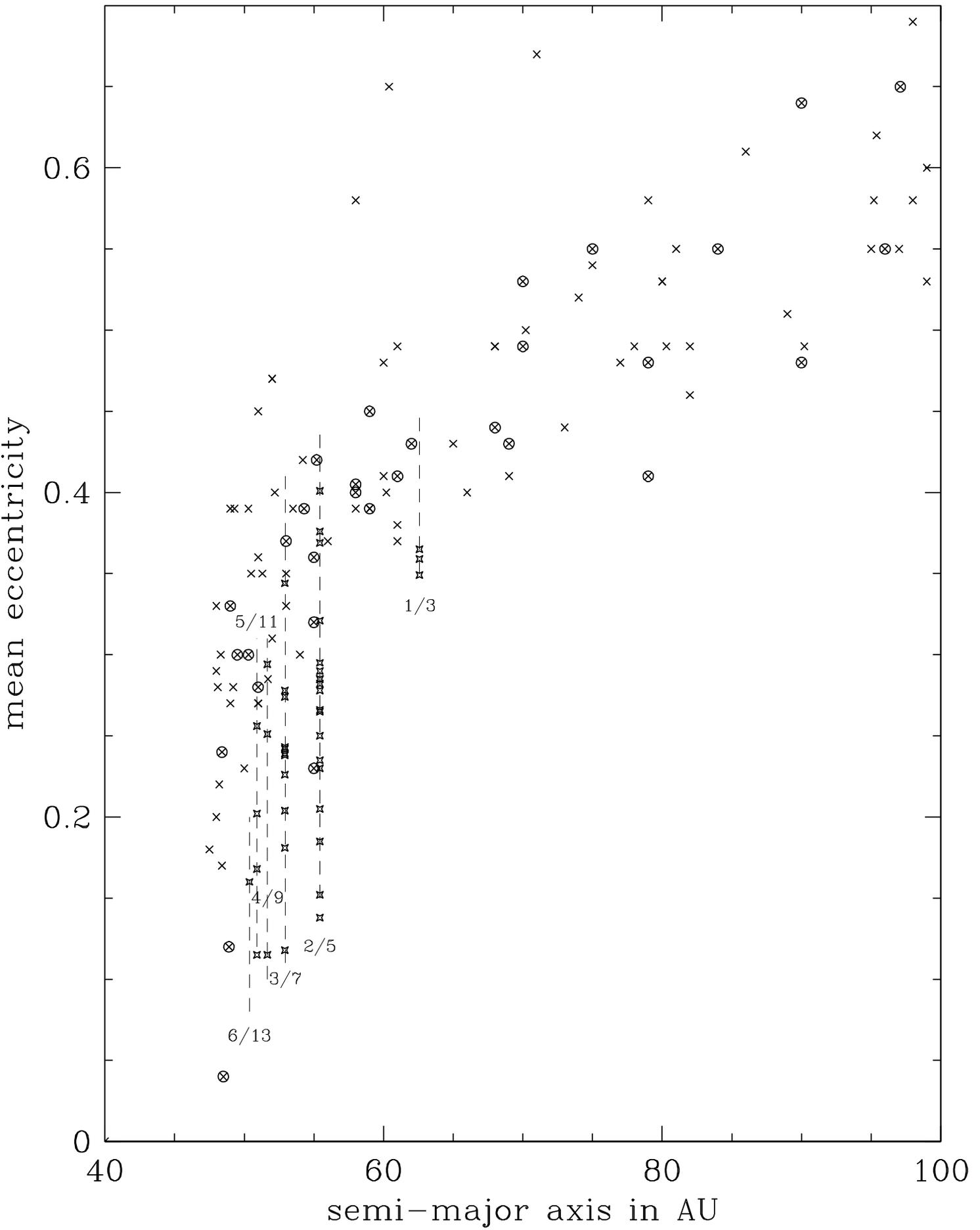

Fig. 1

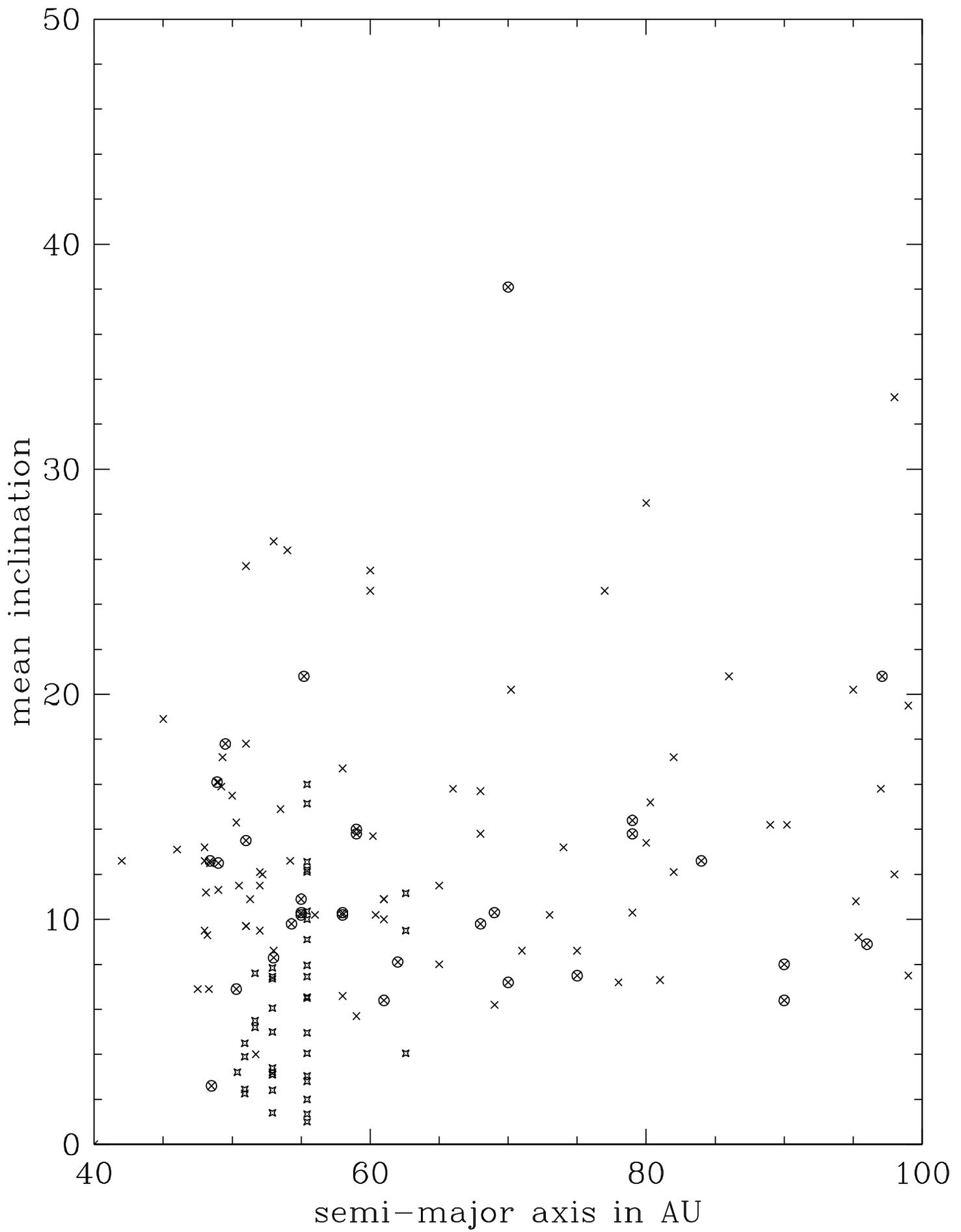

Fig. 2

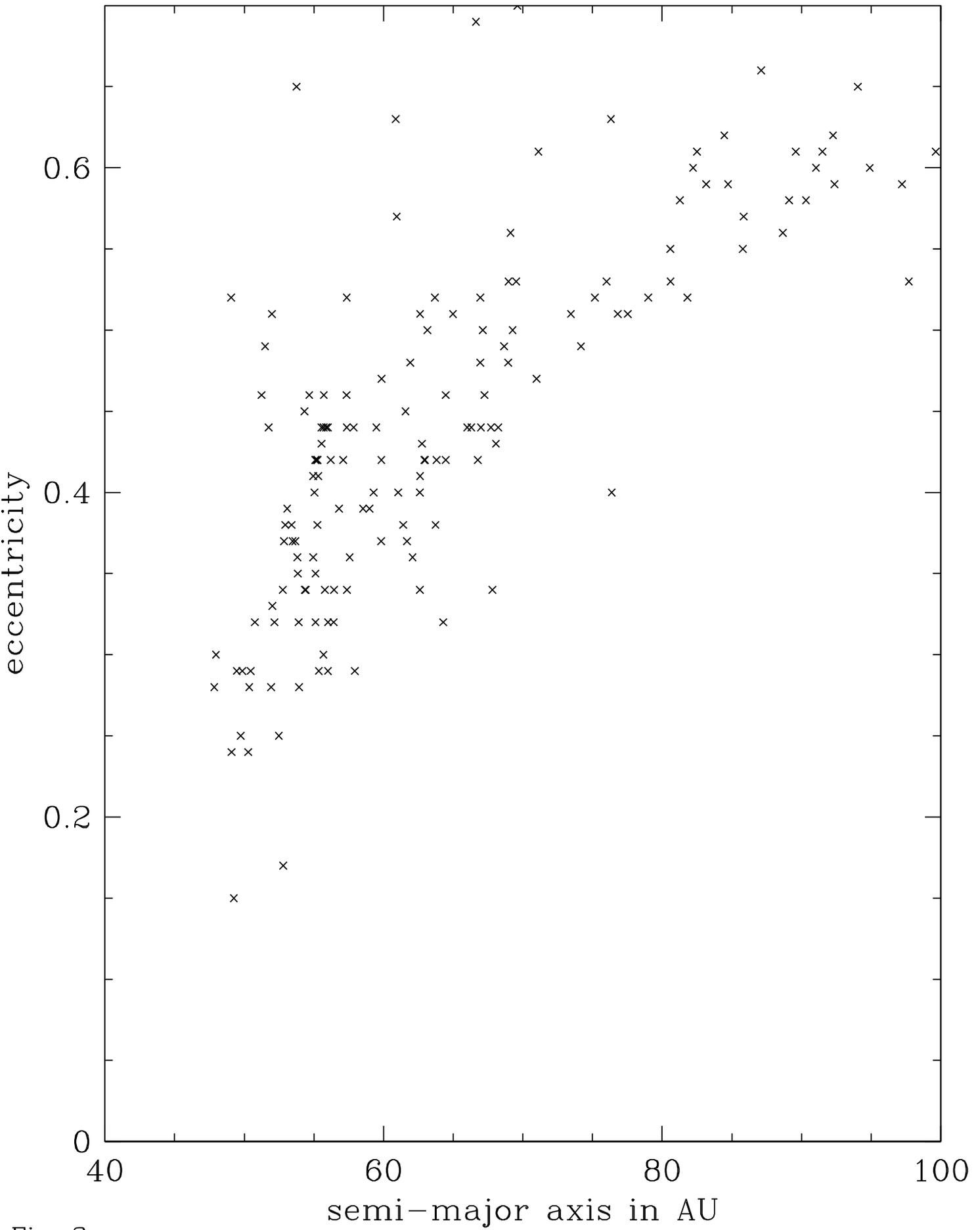

Fig. 3

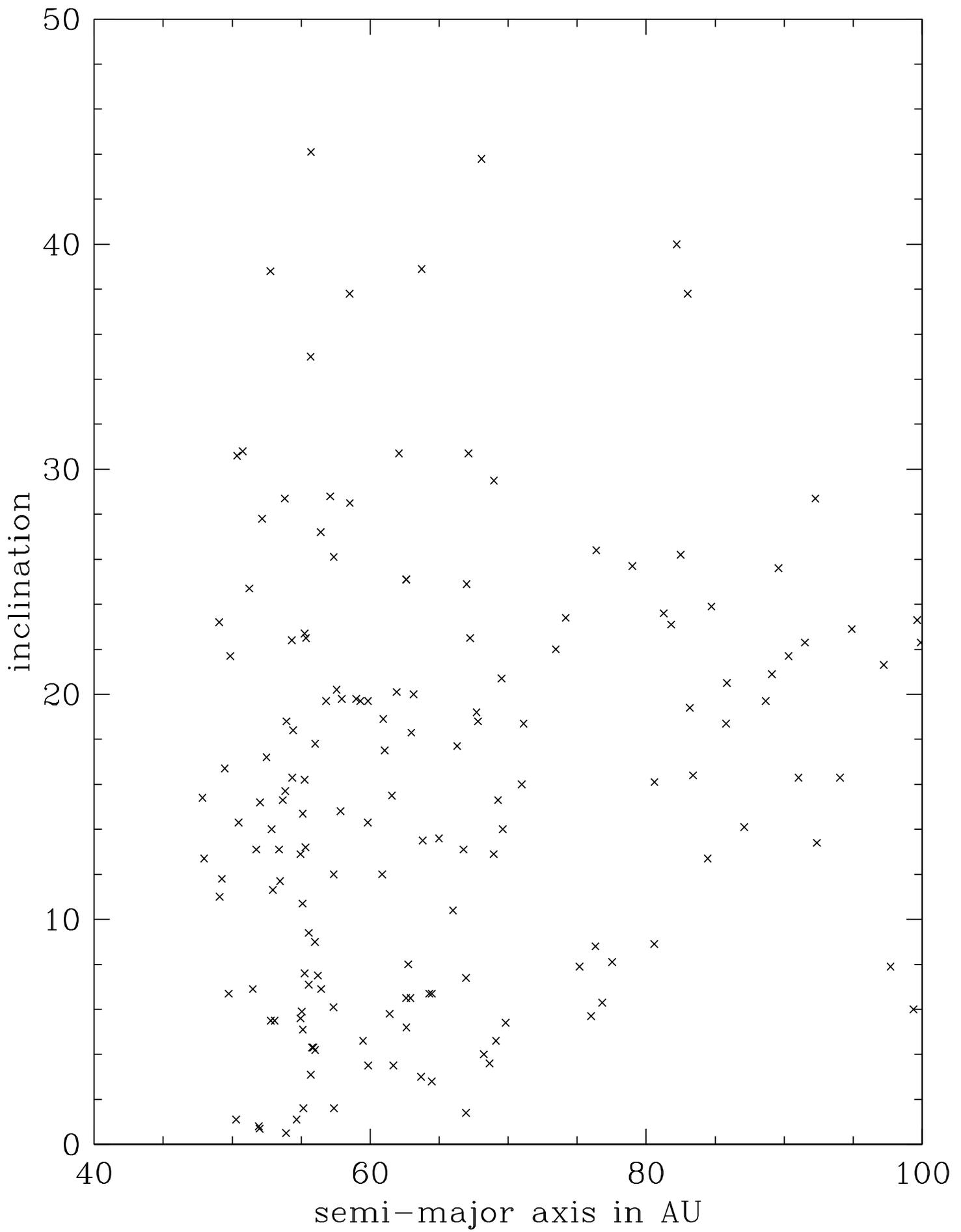

Fig. 4

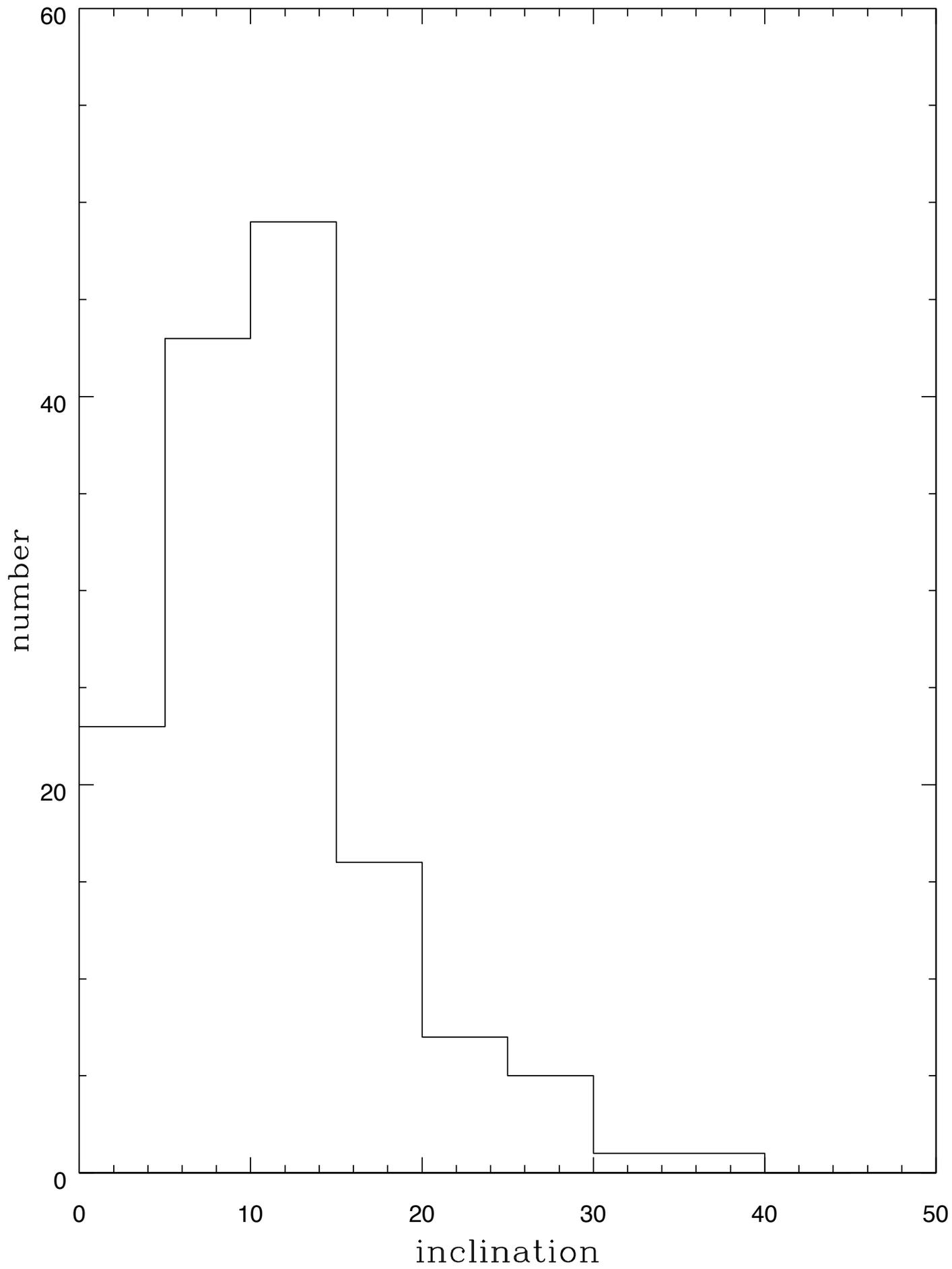

Fig. 5

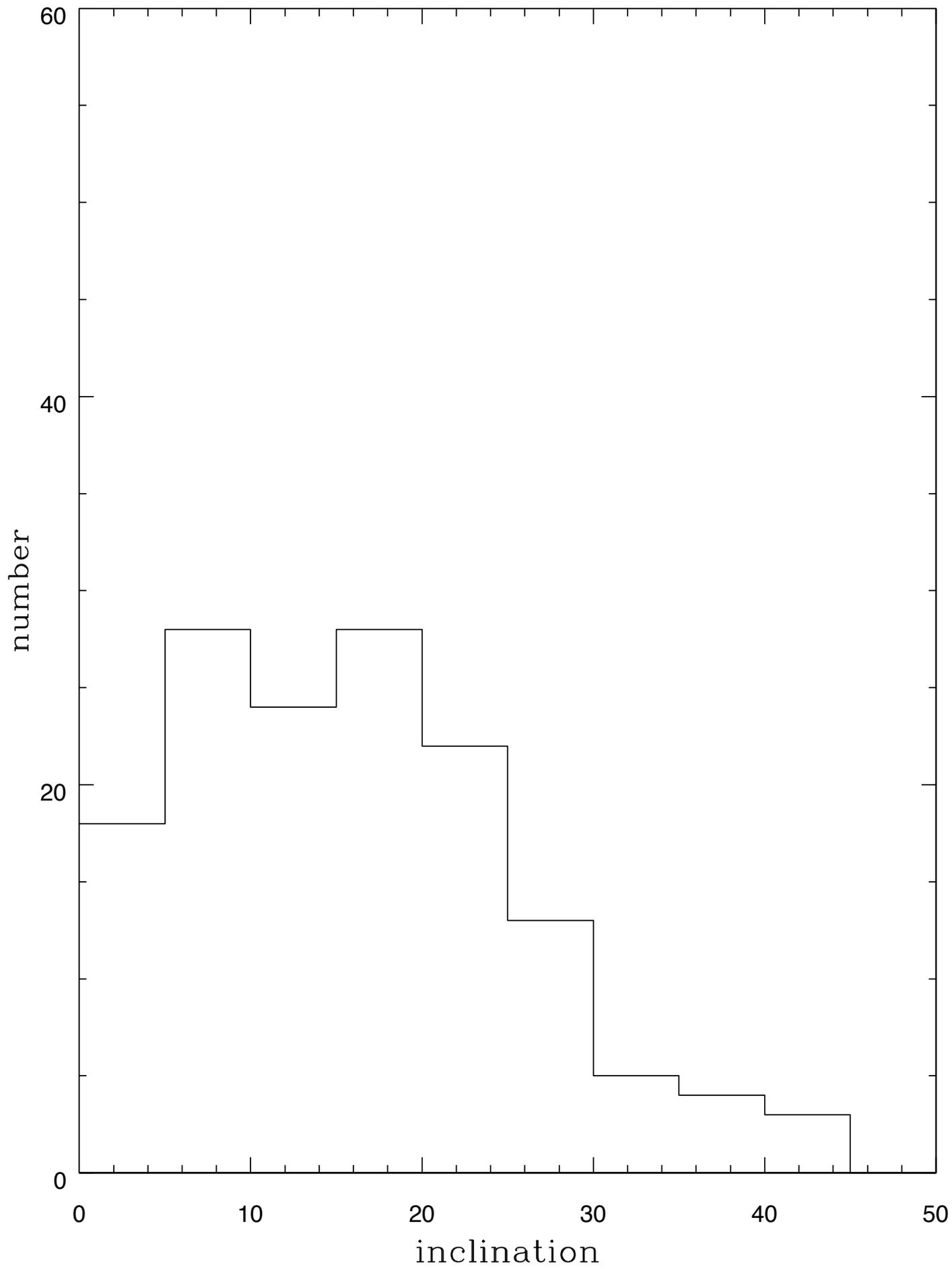

Fig. 6

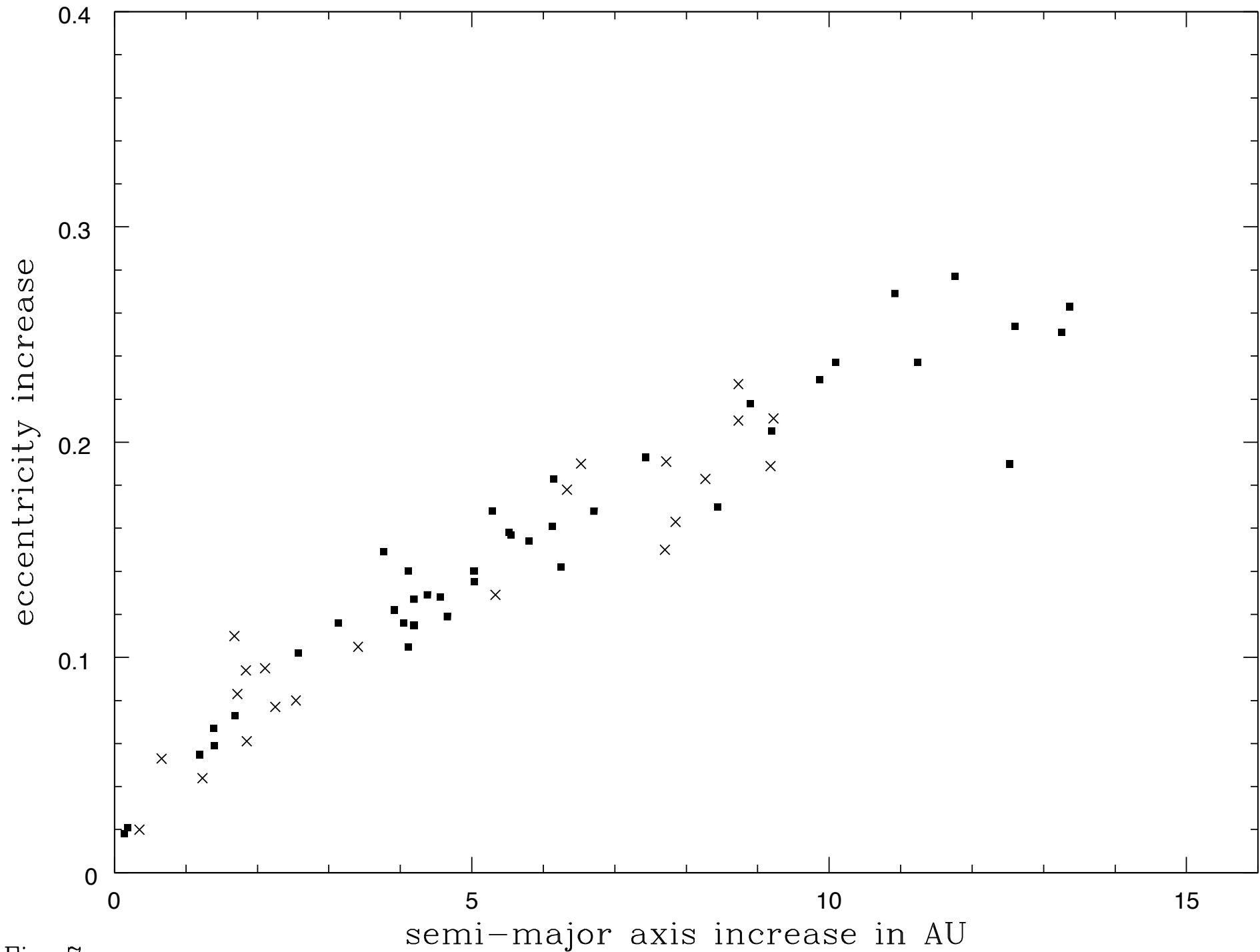
Fig. 7

**Table I. Initial and final locations of mean motion resonances between a(2/3) and a(2/7), of order less than or equal to 7, when Neptune migrates by 7 AU**

| Resonance | a(o) | a(f) | Resonance | a(o) | a(f) |
|---|---|---|---|---|---|
| 2/3 | 30.26 | 39.43 | 6/11 | 34.59 | 45.07 |
| 13/20 | 30.77 | 40.10 | 7/13 | 34.89 | 45.46 |
| 11/17 | 30.86 | 40.22 | 8/15 | 35.11 | 45.75 |
| 9/14 | 31.00 | 40.40 | 1/2 | 36.65 | 47.76 |
| 7/11 | 31.21 | 40.67 | 6/13 | 38.66 | 50.38 |
| 12/19 | 31.37 | 40.88 | 5/11 | 39.06 | 50.90 |
| 5/8 | 31.59 | 41.16 | 4/9 | 39.65 | 51.67 |
| 8/13 | 31.91 | 41.59 | 3/7 | 40.62 | 52.93 |
| 11/18 | 32.06 | 41.78 | 5/12 | 41.39 | 53.94 |
| 3/5 | 32.46 | 42.30 | 2/5 | 42.53 | 55.43 |
| 10/17 | 32.89 | 42.86 | 3/8 | 44.40 | 57.86 |
| 7/12 | 33.07 | 43.10 | 4/11 | 45.32 | 59.06 |
| 4/7 | 33.53 | 43.70 | 1/3 | 48.03 | 62.59 |
| 9/16 | 33.89 | 44.16 | 3/10 | 51.52 | 67.14 |
| 5/9 | 34.17 | 44.53 | 2/7 | 53.23 | 69.36 |

**Table II. Capture probabilities in % at the strongest outer belt resonances**

| Resonance | P(c) 0.04 < e(o) < 0.15 | P(c) e(o) < 0.10 | P(c) e(o) > 0.10 |
|---|---|---|---|
| 1/2 | 22.1 (69) | 31.1 (51) | 12.2 (18) |
| 2/5 | 5.1 (19) | ~3 (5) | 7.7 (14) |
| 3/7 | 2.6 (11) | (2) | 4.6 (9) |
| 1/3 | ~2 (3) | (1) | ~3 (2) |
| others: 4/9, 5/11, 6/13 | (10) (3) (6) (1) | (0) | (10) |

Capture probabilities, P(c), when Neptune migrates by 7 AU over an e-folding time of 10 myr. They are derived from 500 bodies, initially lying in the range: 38 < a(o) < 52 AU., with 0.04 < e(o), sini(o) < 0.15 and are ratios of clearly defined captures to all bodies availably located. In all Tables, the number of bodies is in ( )

**Table III. Capture probabilities, P(c), to 5th order in the outer belt**

| Resonance | P(c) 0.10 < e(o) < 0.15 | | orbits are: chaotic | regular | P(c) 0.15 < e(o) < 0.20 | |
|---|---|---|---|---|---|---|
| 2/7 | | (0) | | | ~1 | (3) |
| 1/3 | 5.0 | (25) | 20 | 5 | 5.0 | (25) |
| 3/8 | ~1 | (3) | 2 | 1 | ~2 | (5) |
| 2/5 | 6.0 | (14) | 11 | 3 | 7.0 | (16) |
| 3/7 | ~4 | (5) | 3 | 2 | ~3 | (4) |
| 4/9 | ~4 | (5) | 3 | 2 | | (1) |
| others: 5/11, 4/11 5/12, 6/13 | | (6) | | | | (6) |

Capture probabilities at resonances from scans with bodies placed in 48.0 < a(o) < 62.6 AU for the same parameters and i(o)'s as the case in Table II, but for two separate runs with the two indicated eccentricity ranges. No escapes from these resonances occurred during 2 byr integrations for the 0.10 < e(o) < 0.15 case, while in the second, shorter calculations provide only P(c).

**Table IV.    Outer belt P(c)'s for T(migration) = 3 myr.**

| Resonance | P(c) 0.04 < e(o) < 0.15 | | P(c) e(o) < 0.10 | | P(c) e(o) > 0.10 | |
|---|---|---|---|---|---|---|
| 1/2 | 18.2 | (58) | 23.2 | (38) | 13.0 | (20) |
| 2/5 | 3.4 | (13) | ~2.5 | (5) | 4.3 | (8) |
| 3/7 | 2.1 | (9) | | (0) | 4.3 | (9) |
| 1/3 | | (0) | | (0) | | (0) |
| others: 4/9, 6/13 | | (5) | | (0) | | (5) |

A case repeating the one of Table II, reducing only the migration time to 3myr from 10 myr.

**Table V.   Outer belt P(c)'s for Neptune's migration of 8AU**

| Resonance | P(c) 0.04 < e(o) < 0.15 | P(c) e(o) < 0.10 | P(c) e(o) > 0.10 |
|---|---|---|---|
| 1/2 | 20.9  (70) | 24.3  (43) | 16.9  (27) |
| 2/5 | 3.9   (16) | ~2    (5)  | 5.8   (11) |
| 3/7 | 1.7   (8)  |       (1)  | 3.3   (7)  |
| 1/3 |       (0)  |       (0)  |       (0)  |
| others: 4/9, 5/11, 6/13 | (9) | (1) | (8) |

A case parallel to that of Table II, but increasing Neptune's migration to 8 from 7AU, with the only other change that the corresponding initial semimajor axis distribution here is from 36 < a(o) < 52 AU.

**Table VI.   Observed population in outer belt resonances**

| Resonance | | Observed Number | Corrected Number |
|---|---|---|---|
| 1/3 | 62.6 AU | 7  | 13 |
| 2/5 | 55.4    | 20 | 20 |
| 3/7 | 52.9    | 9  | 7  |
| 4/9 | 51.7    | 7  | 5  |

Observed number of objects lying within 0.5 AU of the indicated resonance as of June 2012. The only correction for a relative selection effect used here is the 4th power of the semimajor axis.

**Table VII.    Properties of three inner belt [a < a(1/2)] resonances**

| Resonance | P(c) % 0.04 < e(o) < 0.15 | | No. ejected after capture | | remaining orbits are: chaotic | | regular | |
|---|---|---|---|---|---|---|---|---|
| 3/5 | 15.2 | (12) | 4 | [2] | 5 | [4] | 3 | [3] |
| 4/7 | 7.2 | (9) | 6 | [3] | 3 | [2] | 0 | |
| 5/8 | ~10 | (5) | 5 | [1] | 0 | | 0 | |
| 5/9 | ~5 | (5) | 4 | [3] | 1 | [1] | | |

Results here are drawn from the same sample of 500 orbits that provided Table II [and IX], whence T(m) = 10 myr and Neptune's migration is 7 AU. All regular orbits have inclinations less than 8 deg. and average near 5 deg. Numbers in[] here and in other Tables refer to objects with e(o) less than 0.1 and those in ( ) give the total number of bodies.

**Table VIII.   Capture probabilities for three inner belt mmrs in two cases:**

| | a) Neptune migrates outward by 7 AU over 3 myr | | | b) Neptune migrates outward by 8 AU over 10 myr | | |
|---|---|---|---|---|---|---|
| Resonance | P(c) | | | P(c) | | |
| 3/5 | 17.5 | (14) | [11] | 16.5 | (19) | [13] |
| 4/7 | 10.1 | (13) | [5] | 6.8 | (11) | [6] |
| 5/8 | ~12 | (7) | [1] | | (2) | [1] |

In both cases more than 100 bodies lie temporarily in resonances of order four and higher before escaping, often to be captured briefly in other resonances.

**Table IX.    Fate of bodies in Table II after 4.6 billion years**

| Resonance | No. ejected after capture | orbits still in the mmr are: chaotic | | regular | |
|---|---|---|---|---|---|
| 1/2 | 31 [17] | 35 | [31] | 3 | [3] |
| 2/5 | 0 | 16 | [4] | 3 | [1] |
| 3/7 | 0 | 7 | [1] | 4 | [1] |
| 1/3 | 0 | 2 | [0] | 1 | [1] |
| others: 4/9, 5/11, 6/13 | 2 [0] | 3 | [0] | 5 | [0] |

Orbital characteristics of bodies in Table II after 4.6 byr. Column 2 lists the number of captured bodies that have eventually escaped from a resonance, meaning that their semimajor axes lie either in the outer belt, or well beyond it, a > 100 AU. Numbers in brackets refer to bodies with e(o) < 0.1. The two escapers in the final line were both from the 5/11 mmr, ejected at 2.85 and 3.16 byr.